A new multi-metric approach for quantifying global biodiscovery and conservation priorities reveals overlooked hotspots for amphibians.

Sky Button[1] and Amaël Borzée[2,3]

[1]Department of Biology, School of Biological Sciences, Washington State University, Vancouver, WA 98686, United States; 0000-0003-4648-5608; sky.button@wsu.edu

[2] Laboratory of Animal Behaviour and Conservation, College of Biology and the Environment, Nanjing Forestry University, Nanjing, People's Republic of China; 0000-0003-1093-677X; amaelborzee@gmail.com

[3] Jiangsu Agricultural Biodiversity Cultivation and Utilization Research Center, Nanjing, Jiangsu Province, 210014, China

*Correspondence: sky.button@wsu.edu; amaelborzee@gmail.com

**Abstract**

Undocumented species represent one of the largest hurdles for conservation efforts due to the uncertainty they introduce into conservation planning. Until the distribution of earth's biodiversity is better understood, substantial conjecture will continue to be required for protecting species from anthropogenic extinction. Therefore, we developed a novel approach for identifying regions with promising biodiscovery prospects, linked to integrative conservation priorities, which we illustrate using amphibians. Our approach builds on previous estimates of biodiscovery priorities by simultaneously (1) considering linkages between spatio-environmental variables and biodiversity, (2) accounting for the negative relationship between past sampling intensity and future biodiscovery potential, (3) incorporating a priori knowledge about global species distribution patterns, (4) addressing spatial autocorrelation in community composition, and (5) weighting theoretical undocumented species by their predicted levels of conservation need. Using boosted regression trees and 50km$^2$ map pixels spread across the global range of amphibians, we identified several regions likely to contain many undocumented amphibian species and conservation needs, including the Southeast Asian Archipelago, humid portions of sub-Saharan Africa, and undersampled portions of the Amazon, Andes Mountains, and Central America. We also ranked top-scoring ecoregions by their mean and maximum biodiscovery potential and found that the top-20 ranked ecoregions were most concentrated in the Southeast Asian Archipelago and tropical Africa for undocumented species richness, and in tropical Africa and tropical South America for integrative undocumented amphibian conservation needs. However, high-scoring pixels tended to be widely distributed across different ecoregions for both biodiscovery scoring approaches. Our integrative biodiscovery scoring approach is the first to enable the targeting of specific regions for biodiscovery efforts based on explicit, holistic estimates of the conservation importance of such efforts, and therefore represents an important step forward for documenting and preserving earth's biodiversity.

**Introduction**

A key task for conserving organisms in undersampled or overexploited regions is describing species and mapping their distributions prior to their extinction [1]. However, cataloguing new species is a challenging task given that 5–11 million species likely exist on earth [2,3], with the vast majority still undescribed [4]. Moreover, while undescribed species may be vulnerable to extinction due to a lack of knowledge of their existence [1], described species can be similarly vulnerable if their distributions are poorly understood. For instance, development projects have led to the extinction of microendemic taxa that were erroneously assumed to occur more widely [5], whereas more widely-distributed species might also be overlooked in local conservation decision-making if their distributions are underestimated [6]. Alarmingly, the current rate of species extinctions may now exceed 1000x its background rate [4], making it critical to identify and inventory regions with high levels of undocumented biodiversity to improve protected area planning and limit future species extinctions.

Biodiscovery initiatives, which seek to uncover undescribed and locally undocumented species, are a key avenue for bridging biodiversity knowledge gaps and an important first step towards limiting future species extinctions. Unfortunately, due to the limited quantitative guidance available, these initiatives may be forced to choose areas for bioinventory based on unreliable simple heuristics or anecdotal evidence [7], potentially limiting their efficiency in locating undocumented species. In contrast, quantitative approaches like species distribution models (SDMs) are often used to identify overlooked suitable regions for a single or numerous species [8] but are not typically conceptualized for streamlining global biodiscovery efforts. For instance, multi-species occupancy models are increasingly popular for forecasting local species richness [9] but require intensive localized surveys and often pre-defined species pools, making them impractical to implement across global scales and in data-poor regions.

Given clear links between region-specific variables and biodiversity (e.g., more species occur in wet, warm, low-latitude regions than elsewhere), it is surprising that quantitative analyses capitalizing on these linkages in pursuit of biodiscovery are not more common. For example, modeled associations between regional variables and biodiversity in well-sampled regions could be used to generate predictions of global biodiversity patterns, and gaps between predicted and observed regional biodiversity could then be used to infer biodiscovery potential in different regions. Guénard, et al. [7] are perhaps the only to utilize this approach, using it to predict global biodiscoveries for ants. These authors wisely warned of potentially biased estimates of biodiscovery potential that could arise if important variables linked to biodiversity are (un)intentionally omitted from models, or if regions used in model-fitting are non-representative of the global variation in factors linked to biodiversity. Based on other spatial ecology studies, the above challenges could be remedied by including spatial eigenvectors as predictor variables, ensuring that model-fitting pixels are representative of global environments and geographies, and adjusting estimates of biodiscovery potential using sampling intensity [10,11]. To our knowledge, no efforts have been made to address these deficiencies and develop a representative model of biodiscovery priorities based on spatial and environmental variables. Moreover, other recent approaches for estimating biodiscovery potential are possibly problematic or limited due to failing to account for recent sampling intensity and/or the influence of spatio-environmental variables on biodiversity [12,13]. In addition, no attempts have been made (to our knowledge) to estimate the collective conservation needs of locally undocumented species within different regions.

Amphibians are excellent model organisms for developing approaches that pinpoint hotspots of undocumented biodiversity, because their small ranges [14], modest dispersal capabilities [15], and habitat specialization [16] result in more spatially heterogenous patterns of biodiversity than for most other taxa. These traits also lend themselves to species sorting within regions [17], thus, amphibian biodiversity should be highly predictable using environmental and geographic variables. In addition, the niche specialization and inability of amphibians to rapidly shift their habitat use and distributions in response to environmental change makes them disproportionately important as both conservation targets and indicators of environmental quality [18-20]. Identifying areas with high levels of undocumented amphibian biodiversity therefore represents a useful starting point for improving knowledge about both global variation in environmental quality and wildlife conservation needs.

We used amphibians as model organisms for developing a new technique to identify global biodiscovery priority regions. As the consequences of conservation action are larger for some species than others, our approach is designed to predict both undocumented species richness and the total undocumented conservation importance of each region based on the integrative conservation scoring technique developed by Button and Borzée [21]. We used climate, geographic (e.g., spatial eigenvectors), and taxonomic data from globally representative map pixels to predict general amphibian biodiversity patterns, then used this model in combination with observed biodiversity and sampling intensity data to predict amphibian biodiscovery potential across the globe. Our work is based on the important but reasonable assumption that areas with high levels of undocumented amphibian biodiversity should (1) feature a high discrepancy between expected and observed biodiversity and (2) not already be heavily sampled. Our approach is among the first to use environment-biodiversity linkages to predict biodiscovery priorities and builds upon previous approaches [7,12,13] by explicitly accounting for the negative relationship between biodiscovery potential and sampling intensity, incorporating new steps to reduce prediction bias, and weighting species and ecoregions by their relative conservation need. Thus, our approach provides a novel and invaluable resource for researchers interested in bridging conservation knowledge gaps through biodiscovery.

**Methods**

*Quantifying amphibian sampling intensity*

We quantified amphibian sampling intensity across the globe using data from the citizen science platform iNaturalist (www.inaturalist.org) and museum specimens; both downloaded from the Global Biodiversity Information Facility (GBIF; www.gbif.org; DOI: 10.15468/dl.tmv89u; 10.15468/dl.9rhqr2; 10.15468/dl.967m74; 10.15468/dl.cfkdsg; 10.15468/dl.2mu723; 10.15468/dl.wzwr35; 10.15468/dl.mm58xv; 10.15468/dl.zwdxjp; 10.15468/dl.zxhqmh). We selected this dataset, as it is the largest global dataset providing GPS coordinates for amphibians, and because records from other large databases (e.g., VertNet) were mostly redundant with those on GBIF. For both iNaturalist observations and museum specimens, we downloaded all available amphibian records with location data (1600s-present) and assumed that the total number of amphibian observations in an area (regardless of species or year) corresponded with the total sampling intensity in that area. We defined 51,373 hexagonal, 50m$^2$ pixels stretching across the global range of amphibians as "sampling areas" and calculated iNaturalist or museum-based sampling intensity by counting the number of amphibian records in each pixel. The area of 50km$^2$ was deemed adequate as many amphibian species can disperse over such an area over their lifetime [15] yet many do not range across much larger areas [22]. As these two datasets produced pixel-level sampling intensity estimates (Supplementary Figure 1) that were only moderately correlated ($r = 0.47$), we chose to combine iNaturalist and museum records into a single dataset for holistically estimating sampling intensity, used hereafter. This holistic approach yielded sampling intensity estimates that were 94% and 74% correlated with museum and iNaturalist-based sampling intensity respectively, thus its estimates were both holistic and universal.

To account for spatial autocorrelation in amphibian community composition, we converted species observations into a geodesic kernel density raster to estimate effective sampling intensity across a continuous area spanning the global range of amphibians [23]. We used a 224,510 km$^2$ search area (i.e., search radius = 267 km) to create this raster, corresponding with the average range size among all amphibian species with available range shapefiles [23]. To facilitate calculations using sampling intensity and biodiversity predictions, we standardized the distribution of our sampling intensity raster by taking the natural log of each raster pixel (to account for the non-linear nature of species accumulation curves), then converting to a 0-1 scale. This ensured that our initial estimates of biodiscovery potential could be multiplied by 1 minus relative sampling intensity, such that sampling intensity would serve as a penalty for biodiscovery scores in well-sampled pixels without over-inflating biodiscovery scores in unsuitable yet poorly sampled pixels.

*Modeling biodiversity in well-sampled pixels*

We identified linkages between spatio-environmental variables and amphibian biodiversity using an appropriate subset of map pixels (defined below), then projected this model across the globe to determine which pixels were most suitable for amphibian biodiversity. We used boosted regression trees (BRTs) to model the above linkages, based on their high predictive performance and our interest in accurate predictions rather than mechanistic explanations of global biodiversity [24]. To identify suitable pixels to include in model-fitting, we first fit a model for species richness based on spatio-environmental variables, in which we had high a priori confidence that all pixels used to fit the model were sufficiently well-sampled to contain relatively accurate estimates of species richness. We defined pixels as "sufficiently well-sampled" if their level of sampling intensity was at least 3 standard deviations above the global average and/or if they had characteristics known to be unsuitable for amphibian biodiversity in general (i.e., > 55 degrees absolute latitude, < 50mm of annual precipitation, and/or being located on an isolated island > 2,000km from the nearest continental landmass) based on previous research [14]. This approach resulted in an initial model fit to species richness for 12,486 pixels meeting the above criteria, which we then used to predict species richness across all pixels globally. We subtracted observed species richness from predicted species richness to roughly estimate undocumented species richness for each pixel, then modeled undocumented species richness as a general function of predicted species richness and sampling intensity for all pixels globally and re-estimated undocumented species richness as an output of this general function so that we could identify pixels to include in future modeling based on their sampling intensity and initially-predicted undocumented species.

We identified additional pixels to include in model-fitting based on estimates of their sampling intensity and potential for undocumented species (Figure 2). Specifically, we identified 700 pixels that did not meet our initial selection criteria for model-fitting yet featured levels of initially predicted unobserved species and sampling intensity that were > 1 standard deviation below (for unobserved species) and above (for sampling intensity) the respective global average. As a result, these pixels are likely to contain accurate biodiversity estimates and thus were suitable to include in our model-fitting set. Thus, we added these pixels to our initial set of 12,486 model-fitting pixels, resulting in 13,186 pixels used to fit our final biodiversity models, spanning all amphibian-occupied continents and biomes. Despite only increasing the total number of pixels used for model-fitting pixels by 5.6%, this approach more than doubled the number of model-fitting pixels representing four important biomes for amphibians (temperate semi-arid, warm Mediterranean, tropical semi-arid, and tropical humid) [25], while also ensuring that biodiversity estimates within our set of model-fitting pixels remained broadly accurate.

Related to the above, selecting pixels for model-fitting required careful consideration of tradeoffs between the quality of biodiversity estimates used for model-fitting and the spatio-environmental representativeness of an inherently non-random subset of global pixels. Thus, the pixel selection methods described above are further justified hereafter. As amphibians occur in a wide variety of environments and regions, failure to capture this variation during model construction would have reduced our power to predict global biodiversity patterns in a comprehensive manner. A large sample size was therefore required to ensure that pixels used for model-fitting were globally representative. On the other hand, increasing the number of pixels used for model-fitting required relaxing assumptions about how well-sampled each pixel needed to be to support inclusion in the model. Taken to an extreme, fully maximizing sampling size (i.e., using every global pixel to fit models) would have required a large proportion of pixels being included in model-fitting despite insufficient sampling effort to characterize their true biodiversity, resulting in widespread underestimates of biodiversity for pixels with similar spatio-environmental characteristics. This tradeoff was complicated to quantify due to our mixture of categorical and continuous predictor variables. Therefore, we used multiple quantitative and biology-informed qualitative lines of evidence to guide choices about which pixels (and how many) to use to fit models of global biodiversity patterns. For instance, biome was consistently one of the most influential variables in our models regardless of the exact approach used, thus we assumed that using a sufficient number of model-fitting pixels to represent all amphibian-inhabited biomes was important. Secondly, we used well-documented aspects of global amphibian distribution patterns [14] to identify reasonable a priori assumptions: that harsh polar climates (> 55 degrees absolute latitude), extreme deserts (< 50mm of annual precipitation), and islands > 2000km from continental landmasses should have relatively low levels of undocumented amphibian biodiversity (because of low overall biodiversity), regardless of sampling intensity. Thirdly, we used an exploratory modeling approach based on a strict definition for well-sampled pixels (sampling intensity > 3 standard deviations above average, unless the pixel was in an a priori unsuitable location for biodiversity, as defined above) to identify pixels that were unlikely to have high levels of undocumented biodiversity, then included these pixels from future model-fitting if they also had at least moderately high sampling intensity, allowing a balance between the global representativeness and accuracy of biodiversity estimates for our model-fitting pixels. Lastly, we were able to qualitatively validate our final set of model-fitting pixels as more useful than earlier sets tested, because this set resulted in the fewest number of biologically implausible predictions about amphibian biodiversity (e.g., unexpectedly high predictions of amphibian biodiversity in inhospitably cold or dry regions) (Supplementary Figure 6). All of the above factors should be carefully considered when adapting our approach to other taxa or sampling grains to ensure biological plausibility, as model outputs were notably sensitive to choices about which pixels to include in model-fitting.

We initially used biome, continent, mean annual precipitation and temperature (within the focal pixel and within a 1000km radius of the pixel), number of documented amphibian families, approximate average percentage of area within a 1000km radius covered by land (or freshwater; as opposed to saltwater) over evolutionary timescales (i.e., assuming that eustatic sea level was equivalent to its average between the Last Glacial Maximum and Last Interglacial), mean elevation, standard deviation of elevation, elevational range, and two sets of spatial eigenvectors in our BRTs to predict amphibian biodiversity (Table 1). However, the second set of spatial eigenvectors used (which was completely non-redundant with the first set) was more heavily correlated with precipitation (both inside each pixel and within 1000km) for model-fitting pixels than non-model-fitting pixels, potentially causing misleading predictions of biodiversity across the global range of amphibians [26]. To prevent this possibility, we conditioned the two precipitation variables against the above spatial eigenvector set and used this modified version of the precipitation variables in a revised set of our biodiversity models.

We modeled the relationship between the above predictor variables and biodiversity using two sets of boosted regression trees, which predicted (1) species richness or (2) the integrative, conservation-weighted biodiversity score developed by Button and Borzée [21]. To optimize BRT accuracy and consistency as indicated by exploratory analyses, we used settings of tree complexity = 3, learning rate = 0.01, and bag fraction = 0.5 in our final BRT models. After running a set of models that included all possible predictor variables, we discarded each predictor variable that had a relative influence totaling < 5% that of the most influential variable in each model to avoid overfitting, then reran the models with the remaining subset of predictor variables. Importantly, our analyses assumed that amphibian biodiversity had been completely or near-completely described in our model-fitting pixels, hence the detailed process described above for selecting these pixels.

After identifying linkages between spatio-environmental variables and biodiversity within our model-fitting pixels, we used the above models to predict amphibian species richness and integrative conservation priorities [21] across the global range of amphibians. To quantify gaps between expected and observed biodiversity, we subtracted the further from the latter after standardizing (z-scoring) both variables (Figure 2). We refer to this metric as the "theoretical biodiversity gap" of each pixel hereafter.

*Estimating priority areas for biodiscovery*

We operated under the assumption that regions with relatively large amounts of undocumented species and integrative conservation needs should possess high environmental suitability and modest or low sampling intensity. Thus, we calculated global amphibian biodiscovery priorities by multiplying each pixel's theoretical biodiversity gap by its relative scarcity of observations, calculated by standardizing sampling intensity to a 0-1 scale, then subtracting each pixel's standardized sampling intensity from 1. In other words, we used available sampling intensity data to make our "theoretical biodiversity gap" estimates more realistic, reconciling these theoretical estimates with de facto sampling intensities, which are likely to partially explain levels of observed biodiversity in most regions. We used sampling intensity-corrected estimates of biodiversity gaps ("biodiscovery priority scores" hereafter) to map global biodiscovery priorities across the global range of amphibians. We also calculated the mean and maximum priority score within each of 768 amphibian-inhabited ecoregions (among pixels that they occupied a plurality or majority of) [27] from either our species richness or integrative conservation scoring approach [21], then ranked the top 20 highest-priority ecoregions for future biodiscovery efforts based on their biodiscovery priority scores from each approach. We arbitrarily selected 20 ecoregions as this provides a consequential yet digestible amount of information for informing future biodiscovery efforts. We excluded pixels that did not overlap native ranges of amphibians by at least 20% when calculating these rankings, as well as small ecoregions containing fewer than two pixels that satisfied this criterion (for mean ecoregion scores only). We also attempted an analogous calculation of biodiscovery priorities derived from species accumulation curves but ultimately discarded this approach due to its lack of feasibility for poorly-sampled regions (Supplementary Figure 2).

**Results**

Our sampling intensity maps for amphibians incorporated 3,152,283 total amphibian records, including 1,163,683 iNaturalist observations and 1,988,600 preserved specimens. Amphibian sampling intensity varied widely across the world and was mostly concentrated in developed countries. In particular, western Europe, the United States, Australia, the southern Korean Peninsula, Taiwan Island, most of Central America, Ecuador, and western Colombia had the highest densities of amphibian observations (Figure 1). Museum specimens accounted for most of the of amphibian observations in well-sampled regions, except for a few notable regions (e.g., Republic of Korea and the Netherlands) where the usage of iNaturalist has been exceptionally prolific. In less sampled regions, amphibian observations featured a mix of museum specimens and iNaturalist observations, with the respective proportions varying widely among locations (Supplementary Figure 1).

Our final BRT analyses predicted biodiversity with high accuracy (cv correlation = 0.955 for species richness and 0.940 for our integrative conservation score), suggesting appropriateness of these analyses for predicting global biodiscovery priorities. Generally, the most important variable for predicting biodiversity was the number of documented amphibian families.

However, the relative importance of this and other predictor variables varied widely among our two approaches (Table 1; Supplementary Figures 3, 4).

Priority regions for future biodiscovery efforts appeared to vary only modestly depending on whether researchers are seeking to simply document new species, or specifically document new species with high levels of conservation importance (Figure 3). The Southeast Asian Archipelago, humid portions of sub-Saharan Africa, and undersampled portions of the Amazon, Andes Mountains, and Central America were generally considered to have high biodiscovery potential regardless of the scoring metric used. However, predictions of top 20 ranking ecoregions were more variable between the two approaches. Notably, 55% of top-ranked ecoregions (based on maximum pixel scores) were located in the Southeast Asian Archipelago for undocumented species richness, compared to 0% for undocumented integrative conservation priorities, although both approaches nonetheless suggested that at least a few dozen pixels in this region had biodiscovery potentials ranked in the top 1% globally. Rankings of top-20 ecoregions for biodiscovery were also greatly impacted by whether ecoregions were ranked based on their mean or maximum pixel scores. Relatively large ecoregions that spanned multiple countries often varied widely in the biodiscovery priority scores of their pixels, with many ecoregions having modest scores throughout their core, but higher scores near their confluences with other ecoregions. Based on locations of top-scoring pixels within the top-20 ranked ecoregions for biodiscovery potential, countries represented included Indonesia (8 ecoregions), Brazil (4 ecoregions), Venezuela (2 ecoregions), Papua New Guinea (2 ecoregions), Ethiopia (1 ecoregion), Peru (1 ecoregion), Colombia (1 ecoregion), Philippines (1 ecoregion), and Chile (1 ecoregion) for undocumented species richness, and Ethiopia (4 ecoregions), Peru (3 ecoregions), Venezuela (3 ecoregions), Brazil (3 ecoregions), Colombia (2 ecoregions), Bolivia (1 ecoregion), Ecuador (1 ecoregion), Honduras (1 ecoregion), Madagascar (1 ecoregion), Guinea (1 ecoregion), Liberia (1 ecoregion), Cameroon (1 ecoregion), and Angola (1 ecoregion) for undocumented integrative conservation priorities. Brazil, Venezuela, Peru, Colombia, and Ethiopia were the only five countries appearing at least once on both top-20 lists based on maximum pixel scores. Disregarding ecoregions, 36 of the top 50-scoring pixels for undocumented species richness were located in the Southeast Asian Archipelago overall (26 in Borneo), with the other 14 in South America. In contrast, 24 of the overall top 50-scoring pixels for undocumented integrative conservation priorities were located in Africa (19 in Ethiopia), 24 in South America (15 in Peru), and 2 in Central America.

**Discussion**

Our findings suggest that remaining undescribed amphibian diversity is likely to be concentrated within a handful of overlooked regions across the globe, consisting mostly of rural, humid, economically exploited and amphibian rich regions distributed across the tropics. These findings are largely aligned with current and historic social and economic conditions. For example, 9 of the top 10 natural history museums with the most extensive herpetology collections are located in the United States according to GBIF (gbif.org)[28]. Unfortunately, multiple regions likely to support a high diversity of undocumented amphibians are also adjacent to or conterminous with recent wars and/or social upheaval (e.g., Ethiopia, Venezuela, and West Papua), meaning that conservation scientists must carefully weigh personal safety risks during biodiscovery efforts against the urgency of describing and mapping amphibians that may quietly become extinct without conservation action [1,29].

Large discrepancies existed between top ecoregions for biodiscovery potentials identified using mean versus maximum pixel scores, highlighting the need to carefully consider the spatial scale of biodiscovery-oriented fieldwork during early planning phases. For instance, most ecoregions ranked in the top 20 for biodiscovery potential based on mean pixel scores (65% for undocumented species richness; 70% for our integrative score) were not ranked in the top 20 when using maximum pixel scores instead. This finding highlights the tendency for amphibian biodiversity hotspots to occur at fine spatial scales [22], meaning that localized biodiversity is not always representative of ecoregion-wide biodiversity, and vice versa. Thus, we recommend that biodiscovery initiatives not treat all areas within high-ranking ecoregions (based on mean pixel scores) as being of equal priority, but instead focus on identifying and surveying areas *within* ecoregions that are likely to yield the most undocumented species (e.g., ecotones, moist microclimates, and area with high topoclimate diversity). Our ecoregion rankings based on maximum pixel scores provide a starting point for identifying specific survey regions (Tables 2, 3), but finer-scale analyses are likely needed to elucidate optimal field sites, especially given the time and resource-limited nature of most biodiscovery initiatives [30]. Moreover, several ecoregions with pixels scoring in the top 1% globally for biodiscovery potential nonetheless did not appear in our top 20 lists for any scoring approach. Thus, we recommend not ruling out ecoregions as biodiscovery targets if they are not included on these lists, and instead refer readers to our maps of biodiscovery potential (Figure 3) and a supplemental database listing the scores and geographic centroids of each pixel (available upon request) for additional guidance on amphibian biodiscovery.

Our two scoring approaches yielded broadly similar inferences about global biodiscovery priorities for amphibians, although top-ranking ecoregions varied widely between approaches (Supplementary Figure 5), emphasizing distinctions between the two. Both approaches generally favored the Southeast Asian Archipelago, humid portions of sub-Saharan Africa, and undersampled portions of the Amazon, Andes Mountains, and Central America as having exceptional biodiscovery potential. However, we found notable differences between the top 20 ranking ecoregions for biodiscovery potential for species richness versus the integrative conservation scoring metric developed by Button and Borzée [21]. For instance, 55% of all top-ranked ecoregions ($n = 11$ of 20) based on maximum pixel scores were located in the Southeast Asian Archipelago for undocumented species richness, compared to 0% ($n = 0$ of 20) for undocumented integrative conservation priorities. This finding may be explained by the fact that the Southeast Asian Archipelago may possess less evolutionarily unique amphibians compared to elsewhere in the tropics (e.g., no native salamanders and only a single, relatively diversified genus of caecilians), and by the stable conservation status of most of its top-ranking ecoregions. In addition, the difficulty for species to disperse over the Wallace's Line has likely prevented the establishment and diversification of even more species in this region [31]. Moreover, rates of local endemism on some islands in the Southeast Asian Archipelago (e.g., New Guinea) may be limited due to the relatively flat topography within some of their top-scoring ecoregions for species richness [32,33]. This finding highlights the need for biodiscovery initiatives to not only prioritize new species descriptions and range expansions, but also consider the potential levels of endemism, taxonomic uniqueness, and degree of human threats to undescribed species when selecting target regions for fieldwork [21,22,34]. However, the Southeast Asian Archipelago in particular possessed several pixels with scores in the top 1% globally, including pixels in all its constituent countries regardless of the scoring metric used, making it an unambiguously important region for future biodiscovery efforts. Moreover, we do not suggest that there is no merit to carrying out biodiscovery work in undersampled (or even well-sampled) regions if they do not have a high expected integrative conservation score. Instead, we merely argue that a conservation motive underlying biodiscovery efforts can affect which areas should be most urgently prioritized for future surveys.

In addition to our final approach for predicting biodiscovery priority regions, we also attempted an alternative approach to identify such regions based on region-specific species accumulation curves, which we ultimately discarded (Supplementary Figure 2). Unfortunately, this approach was only viable for regions that had been at least modestly sampled; otherwise, there were not sufficient observations to fit a species accumulation curve to the area. Thus, biodiscovery priorities could not be directly estimated from poorly-sampled regions using this approach, and instead had to be derived by interpolating results from the nearest better-sampled regions. This coarse spatial resolution and interpolation led to clearly unreasonable suggestions about amphibian biodiscovery priorities (e.g., parts of the Sahara Desert were suggested as top priorities). Moreover, biodiversity estimates derived from species accumulation curves were highly uncertain for modestly sampled pixels. Therefore, we do not recommend this method for determining biodiscovery priorities outside of regions that have already been moderately or heavily sampled.

One limitation of our analyses was the difficulty of identifying a set of variables that were universally useful for predicting amphibian biodiversity. For instance, the proportion of the area within 1000km of each pixel that was covered by land was included as a predictor variable to avoid overpredicting biodiversity for humid tropical islands that are environmentally suitable but not colonizable for amphibians. However, some isthmuses (e.g., in Costa Rica and Panama) are colonizable for amphibians yet have a relatively low percentage of land within 1000km, reducing the utility of this predictor variable. Moreover, simply reducing the search radius when calculating this and other analogous variables (e.g., using land within 500km instead of 1000km) was not a feasible solution, as the amount of scale reduction necessary to eliminate the above problem would have created a new problem of the spatial scale being too small to adequately reflect each pixel's colonization potential over long evolutionary timescales. Nevertheless, high cross-validated correlations in our models and a lack of obvious false-positive regions incorrectly highlighted as "high priority" for biodiscovery potential suggested that our final set of predictor variables was relatively accurate in predicting true levels of biodiversity, assuming limited false negatives. Moreover, our results are consistent with broad geographic patterns documented for amphibian biodiversity [14] but provide an enhanced resolution regarding which areas may contain much greater biodiversity than has currently been described. In contrast, earlier versions of our BRTs tended to incorrectly predict high biodiversity in humid temperate regions with oceanic climates at moderate-to-high latitudes (e.g., coastal British Columbia), likely because climatic conditions in these regions were similar to those found in biodiverse, high-elevation cloud forests in the tropics. Including the number of amphibian families documented in each pixel as a predictor variable was invaluable for overcoming this challenge. While we recognize that some amphibian families have not yet been documented in undersampled regions, we deemed it ok to include this variable in our models due to the tendency for new families to be described very infrequently (compared to genera and species), including in undersampled regions. Moreover, no new amphibian families have been described globally since Nasikabatrachidae 20 years ago [35], except for taxonomic revisions based on updated genetics.

Unfortunately, most countries likely to possess high amphibian biodiversity are also overexploited due their rich natural resources; frequently by foreign investors who benefit financially if local populations remain undereducated and thus serve as cheap sources of labor [36]. As a result, careers in biology are rarely economically viable for locals living in the most undersampled and biodiverse regions. However, locals are often more personally familiar with the organisms around them than unacquainted foreign scientists, making it important that more efforts be made to train and collaborate with residents of undersampled countries when conducting biodiscovery research. Additionally, research conducted in underdeveloped countries by foreign scientists should seek to advance the careers of local academics through collaboration and co-authorship. Moreover, foreign scientists should also consider depositing type specimens of new species in institutions accessible to local collaborators [37]. Arguably, one of the largest roadblocks to documenting biodiversity in underdeveloped countries is a savior mentality present in some Western academics that leads to "parachute science" [38,39] and a failure to recognize and support the ability of locals in these countries to succeed in scientific careers. In contrast, research in isolation is not beneficial for maximizing biodiscoveries either [40], and maintaining international collaboration may help alleviate the risk of species loss.

The purpose of this study was to identify general regions that conservation-minded biodiscovery initiatives should consider targeting for future surveys. Our analyses revealed understudied potential amphibian hotspots clustered most densely within a few key regions, such as the Southeast Asian Archipelago, humid portions of sub-Saharan Africa, and undersampled portions of the Amazon, Andes Mountains, and Central America (Tables 2, 3; Figure 3). Due to the coarse resolution of our pixels, finer-scale analyses should be conducted within top priority regions to identify precise locations to sample. For example, genus or family-level species distribution models (SDMs) for taxa located adjacent to (but not yet documented in) priority regions may help elucidate specific best areas to search within these regions to uncover undescribed related taxa and/or document range extensions. In addition, genus- and family-level eDNA sampling in priority biodiscovery regions could help determine the taxonomy of undocumented or "lost" species [41], providing clues about their life histories and which habitats to search to find individuals. Our findings should therefore be treated as a starting point for improving the manner in which field sites for biodiscovery research are chosen and eventually surveyed to uncover undocumented taxa.

**Acknowledgment**

We are thankful to the reviewers for their time spent on our manuscript.

**Funding**

This work was partially supported by the Foreign Youth Talent Program (QN2021014013L) from the Ministry of Science and Technology of the People's Republic of China to AB.

**Statement of conflict**

None of the authors has conflicts of interest to report.

**Tables**
**Table 1.** Variables used for modeling global amphibian biodiversity, and their relative contributions in our final BRTs for species richness and total integrative conservation priority scores. Mean values and ranges are relative to all pixels within the total global range of amphibians.

| Variable | Mean | Range | Species richness contribution | Integrative score contribution |
|---|---|---|---|---|
| Average elevation (m) | 569.3 | -56.1 – 5,582 | — | 2.9% |
| Average annual precipitation (mm) | 902.2 | 0 – 6,901 | — | 4.6% |
| Average annual precipitation within 1000km (mm) | 873.5 | 29.1 – 3,243 | — | — |
| Average annual temperature (°C) | 13.24 | -17.25 – 30.56 | — | — |
| Average annual temperature within 1000km (°C) | 13.06 | -14.53 – 27.90 | — | 4.7% |
| Biome | NA | NA | 13.4% | 19.2% |
| Continent | NA | NA | — | 3.7% |
| Current proportion of area within 1000km covered by land | 0.74 | 0 – 1 | — | — |
| Historical average proportion of area within 1000km covered by land | 0.81 | 0 – 1 | — | 3.8% |
| Maximum minus minimum elevation | 477.8 | 0 – 6,101 | — | 9.2% |
| Number of amphibian families documented | 5.26 | 1 – 18 | 77.5% | 40.8% |
| Spatial eigenvector 1 | 0.00 | -0.08 – 0.04 | — | — |
| Spatial eigenvector 2 | -2.14 | -9,999 – 0.03 | 9.1% | 7.3% |
| Standard deviation of elevation (m) | 109.7 | 0 – 2,049 | — | 3.5% |

**Table 2.** Top 20 ecoregions for future biodiscovery potential based on projected undiscovered species richness. * = unanimous top 20 ecoregion for both mean and maximum undocumented species richness; ** = unanimous top 10 ecoregion (as above); *** = unanimous top 5 ecoregion (as above). Bolded and underlined ecoregions are ranked, respectively, in either the top 10 or 20 ecoregions for their respective metric (mean or maximum priority score), regardless of the scoring approach used (i.e., for undocumented species richness and total undocumented integrative conservation priorities). Countries are listed for top-scoring pixels within ecoregions, but not for ecoregion-wide (mean) scores, as ecoregions often spanned multiple countries.

| Mean Species Richness Priority Score | | | Maximum Species Richness Priority Score | | | |
|---|---|---|---|---|---|---|
| Rank | Ecoregion Name | Continent | Rank | Ecoregion Name | Continent | Country |
| 1 | Admiralty Islands lowland rain forests* | Oceania | 1 | **Peruvian Yungas** | South America | Peru |
| 2 | Southern New Guinea lowland rain forests* | Oceania | 2 | Purus varzeá | South America | Colombia |
| 3 | Southern New Guinea freshwater swamp forests | Oceania | 3 | Sundaland heath forests* | Asia | Indonesia |
| 4 | **Ethiopian montane moorlands*** | Africa | 4 | Borneo peat swamp forests | Asia | Indonesia |
| 5 | **Cross-Niger transition forests** | Africa | 5 | Venezuelan Andes montane forests | South America | Venezuela |
| 6 | **Ethiopian montane grasslands and woodlands** | Africa | 6 | Borneo lowland rain forests | Asia | Indonesia |
| 7 | Biak-Numfoor rain forests | Oceania | 7 | Sulu Archipelago rain forests* | Asia | Philippines |
| 8 | Southern Zanzibar-Inhambane coastal forest mosaic | Africa | 8 | Valdivian temperate forests | South America | Chile |
| 9 | Marajó varzeá* | South America | 9 | Borneo montane rain forests | Asia | Indonesia |
| 10 | Greater Negros-Panay rain forests | Asia | 10 | Pantepui | South America | Venezuela/ Brazil |
| 11 | Tapajós-Xingu moist forests | South America | 11 | New Britain-Latangai Island lowland rain forests | Oceania | Papua New Guinea |
| 12 | Trans Fly savanna and grasslands* | Oceania | 12 | Southern New Guinea lowland rain forests* | Oceania | Indonesia |
| 13 | Zambezian coastal flooded savanna | Africa | 13 | Trans Fly savanna and grasslands* | Oceania | Indonesia |
| 14 | Sundaland heath forests* | Asia | 14 | Uatuma-Trombetas moist forests | South America | Brazil |
| 15 | Miskito pine forests | North America | 15 | Ethiopian montane moorlands* | Africa | Ethiopia |
| 16 | Sulu Archipelago rain forests* | Asia | 16 | Northern New Guinea montane rain forests | Oceania | Indonesia |
| 17 | Ethiopian montane forests | Africa | 17 | Marajó varzeá* | South America | Brazil |

| 18 | Nigerian lowland forests | Africa | 18 | Manus (Admiralty) Islands lowland rain forests* | Oceania | Papua New Guinea |
| 19 | Western Guinean lowland forests | Africa | 19 | Tocantins/Pindare moist forests | South America | Brazil |
| 20 | Isthmian-Pacific moist forests | North America | 20 | Southern New Guinea freshwater swamp forests | Oceania | Indonesia |

**Table 3.** Top 20 ecoregions for future biodiscovery potential based on total undocumented integrative conservation priorities [21]. * = unanimous top 20 ecoregion for both mean and maximum undocumented species richness; ** = unanimous top 10 ecoregion (as above); *** = unanimous top 5 ecoregion (as above). Bolded and underlined ecoregions are ranked, respectively, in either the top 10 or 20 ecoregions for their respective metric (mean or maximum priority score), regardless of the scoring approach used (i.e., for undocumented species richness and total undocumented integrative conservation priorities). Countries are listed for top-scoring pixels within ecoregions, but not for ecoregion-wide (mean) scores, as ecoregions often spanned multiple countries.

| Mean Integrated Priority Score | | | Maximum Integrated Priority Score | | | |
|---|---|---|---|---|---|---|
| Rank | Ecoregion Name | Continent | Rank | Ecoregion Name | Continent | Country |
| 1 | Hengduan Mountains subalpine conifer forests | Africa | 1 | Ucayali moist forests** | South America | Peru |
| 2 | **Ethiopian montane grasslands and woodlands**** | Africa | 2 | **Peruvian Yungas** | South America | Peru |
| 3 | <u>Ethiopian montane forests</u> | Africa | 3 | Southwest Amazon moist forests | South America | Peru |
| 4 | **Cross-Niger transition forests** | Africa | 4 | Somali Acacia-Commiphora bushlands and thickets | Africa | Ethiopia |
| 5 | Jos Plateau forest-grassland mosaic | Africa | 5 | Bolivian Yungas** | South America | Bolivia |
| 6 | Buru rain forests | Asia | 6 | Madagascar subhumid forests | Africa | Madagascar |
| 7 | **Ethiopian montane moorlands*** | Africa | 7 | Ethiopian montane grasslands and woodlands** | Africa | Ethiopia |
| 8 | Rwenzori-Virunga montane moorlands | Africa | 8 | Victoria Basin forest-savanna mosaic* | Africa | Ethiopia |
| 9 | Ucayali moist forests** | South America | 9 | Patía Valley dry forests | South America | Colombia |
| 10 | Bolivian Yungas** | South America | 10 | Negro-Branco moist forests | South America | Brazil |
| 11 | Victoria Basin forest-savanna mosaic* | Africa | 11 | <u>Venezuelan Andes montane forests</u> | South America | Venezuela |
| 12 | Niger Delta swamp forests | Africa | 12 | <u>Ethiopian montane moorlands*</u> | Africa | Ethiopia |
| 13 | Northern Congolian forest-savanna mosaic | Africa | 13 | Western Guinean lowland forests | Africa | Guinea/Liberia |
| 14 | Central American dry forests | North America | 14 | Central American pine-oak forests* | North America | Honduras |
| 15 | Central American Atlantic moist forests | North America | 15 | Napo moist forests | South America | Colombia |
| 16 | Guinean montane forests | Africa | 16 | Mount Cameroon and Bioko montane forests | Africa | Cameroon |
| 17 | Guianan freshwater swamp forests | South America | 17 | Northern Andean páramo | South America | Ecuador |

| 18 | Central American pine-oak forests* | North America | 18 | Angolan montane forest-grassland mosaic | Africa | Angola |
| 19 | Tocantins/Pindare moist forests | South America | 19 | Pantepui | South America | Brazil/Venezuela |
| 20 | Northeastern Congolian lowland forests | Africa | 20 | Guianan piedmont and lowland moist forests | South America | Brazil/Venezuela |

**Figures**

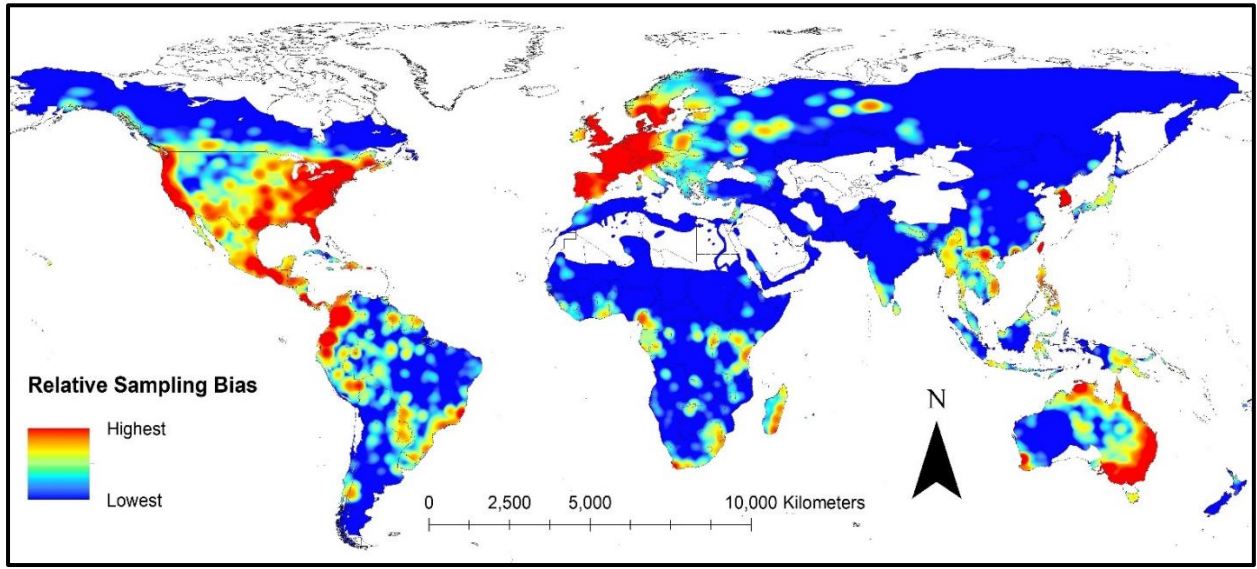

**Figure 1.** Kernel density-based estimates of effective relative sampling intensity across the global range of amphibians, based on combined iNaturalist observations and museum specimens, both downloaded from GBIF. The dataset was downloaded in segments due to its large size and includes the following DOIs: 10.15468/dl.tmv89u; 10.15468/dl.9rhqr2; 10.15468/dl.967m74; 10.15468/dl.cfkdsg; 10.15468/dl.2mu723; 10.15468/dl.wzwr35; 10.15468/dl.mm58xv; 10.15468/dl.zwdxjp; 10.15468/dl.zxhqmh.

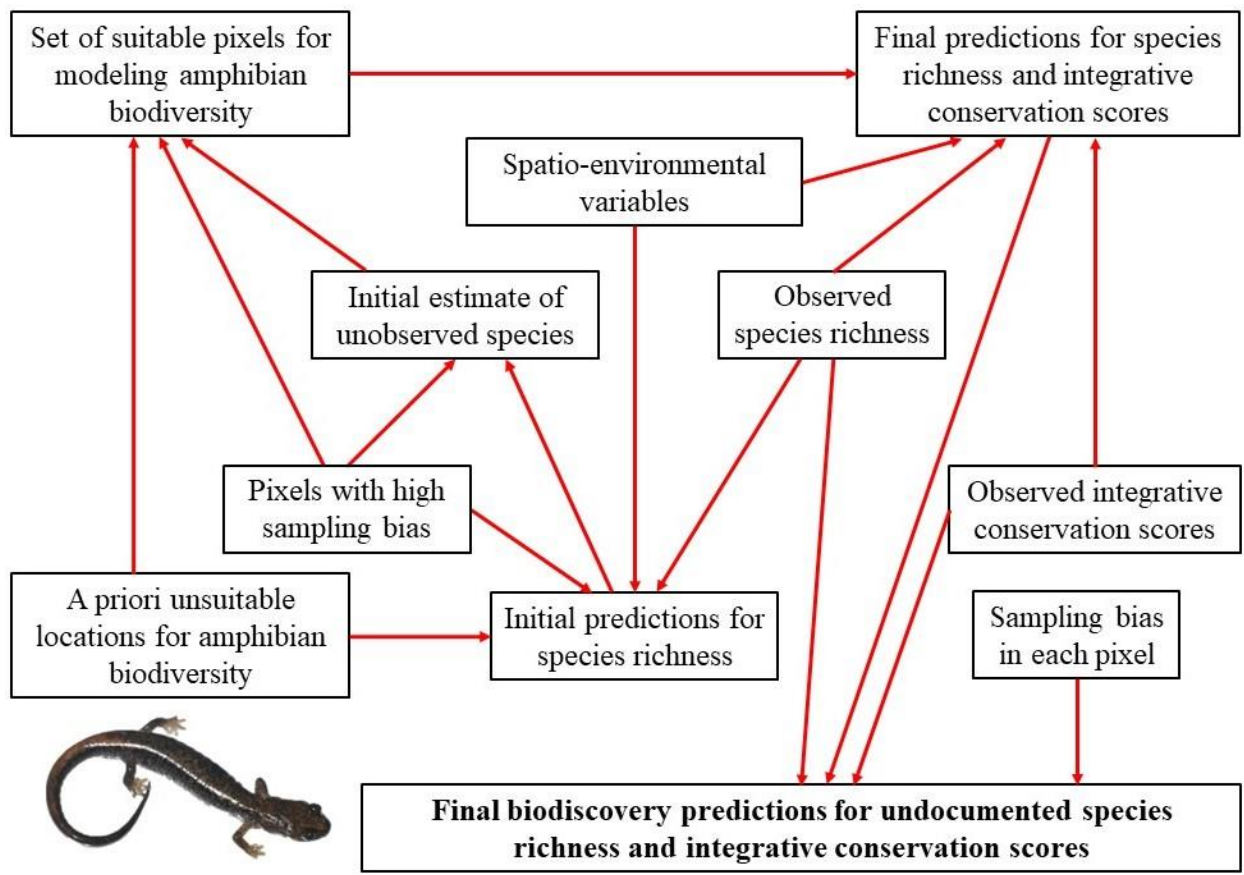

**Figure 2.** A diagram of our approach for estimating global biodiscovery priorities (i.e., "theoretical biodiversity gaps") and identifying suitable pixels for doing so. The representative amphibian species in the picture is *Karsenia koreana*.

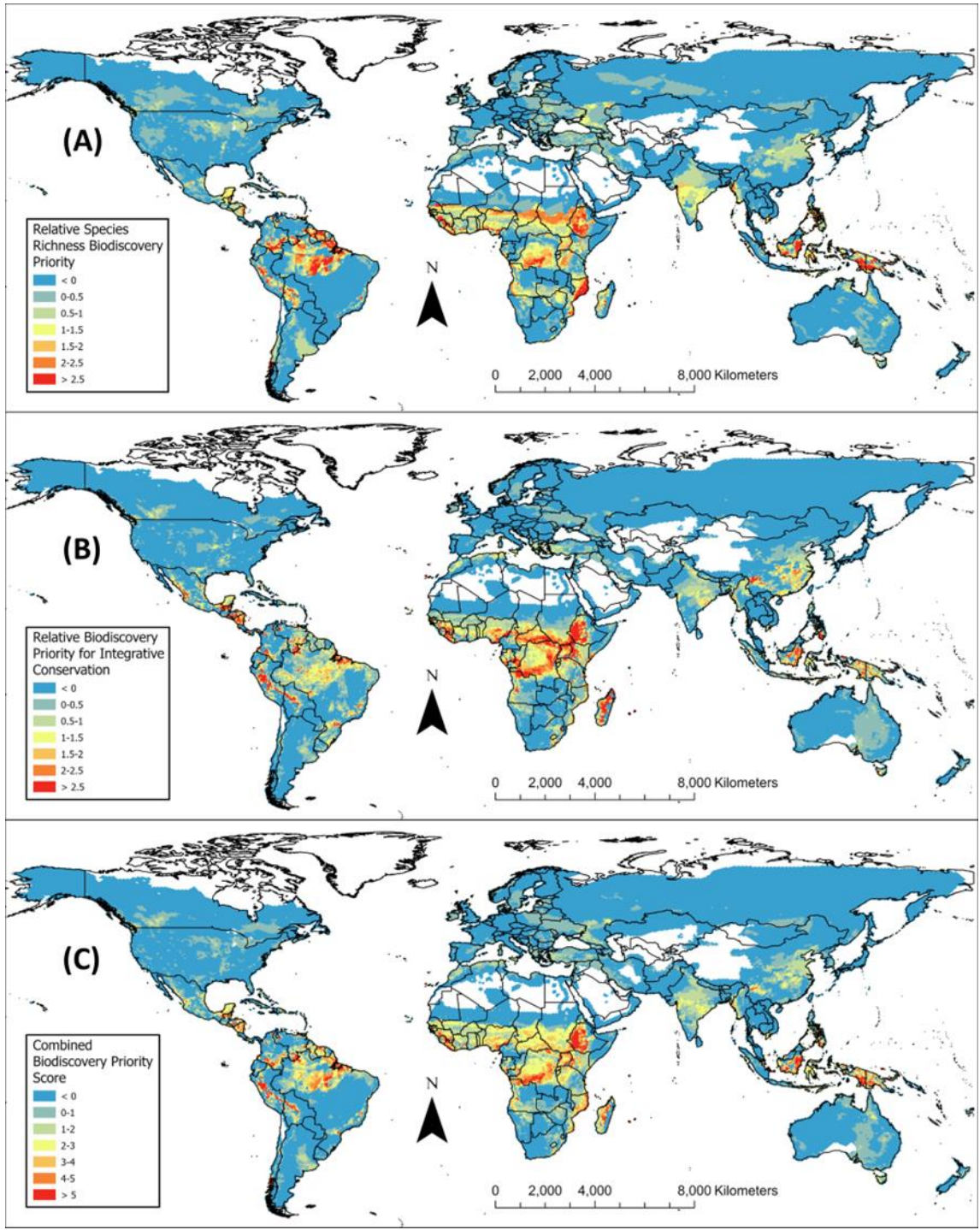

**Figure 3.** Relative global biodiscovery priorities for amphibians based on predicted undocumented species richness (A), total undocumented integrative conservation priorities (B), and a combination (sum) of the two scoring approaches (C).

**Supplementary Data**

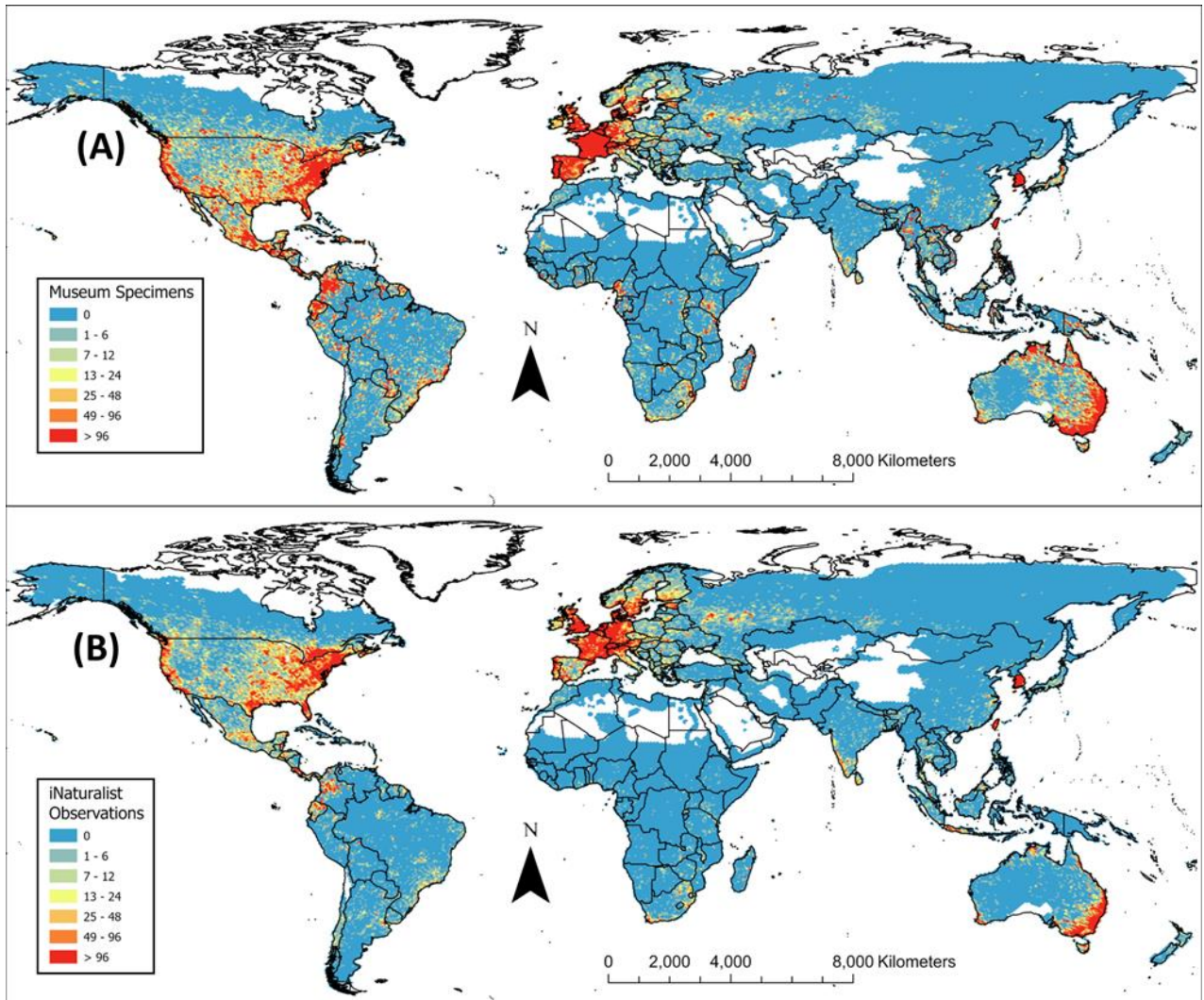

**Supplementary Figure 1.** Global patterns of sampling intensity, based on amphibian museum specimens (A) and iNaturalist observations (B) within each pixel. Datasets downloaded from the Global Biodiversity Information Facility (GBIF; www.gbif.org; DOI: 10.15468/dl.tmv89u; 10.15468/dl.9rhqr2; 10.15468/dl.967m74; 10.15468/dl.cfkdsg; 10.15468/dl.2mu723; 10.15468/dl.wzwr35; 10.15468/dl.mm58xv; 10.15468/dl.zwdxjp; 10.15468/dl.zxhqmh).

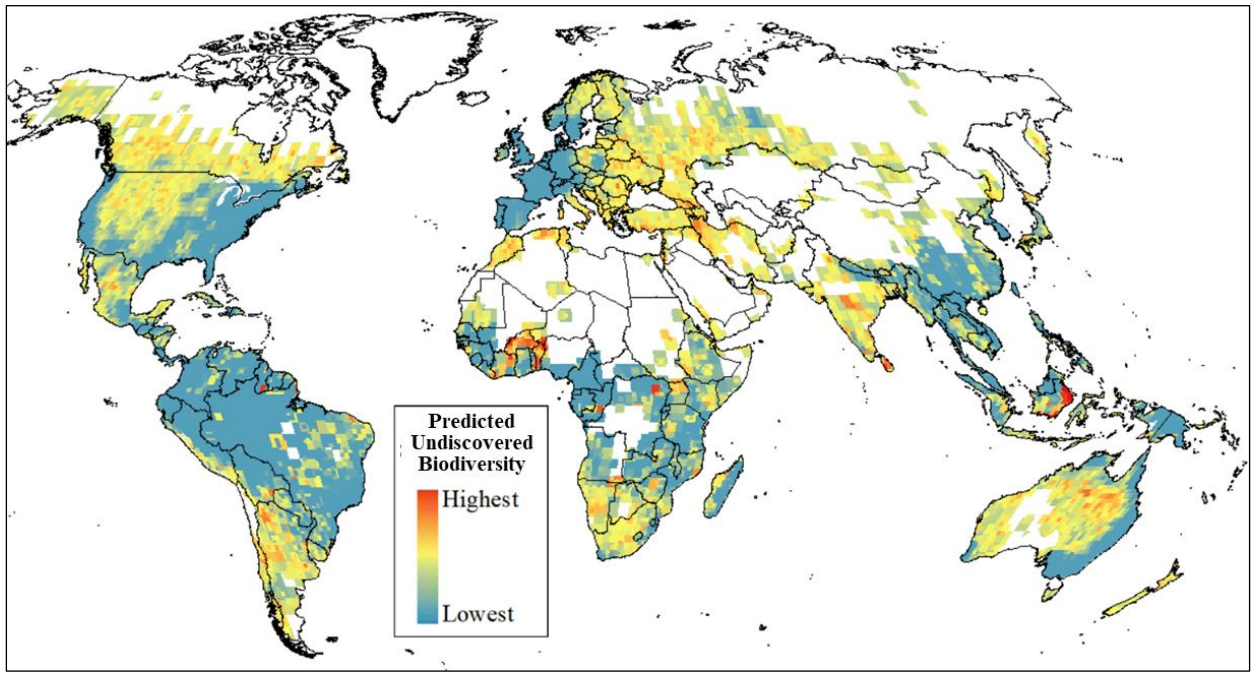

**Supplementary Figure 2.** Map of biodiscovery priorities for expected undescribed species richness, derived by fitting semi-log species accumulation curves to observations within 1x1, 2x2 or 3x3 degree pixels (whichever was the smallest resolution that allowed for curve fitting) and then predicting the number of species observed given sufficient sampling to raise the pixel in question to the 90$^{th}$ percentile for global sampling intensity. To estimate biodiscovery priorities, we subtracted the current known biodiversity of the pixel from our estimate of theoretical biodiversity in the pixel expected given adequate sampling. Blue pixels represent the lowest priority areas, while red pixels represent the highest. We ultimately discarded this approach, because there were too few amphibian observations in the most undersampled regions (which potentially contain the most undescribed species) to map biodiscovery priorities for these regions, hence the white spaces on the map. Moreover, it was unclear how heavily sampled pixels needed to be to produce informative species accumulation curves and thus reasonable estimates of biodiscovery potential, which may explain unintuitive results depicted on this map for some cold and/or dry regions (e.g., Canada, the Australian Outback, and Burkina Faso).

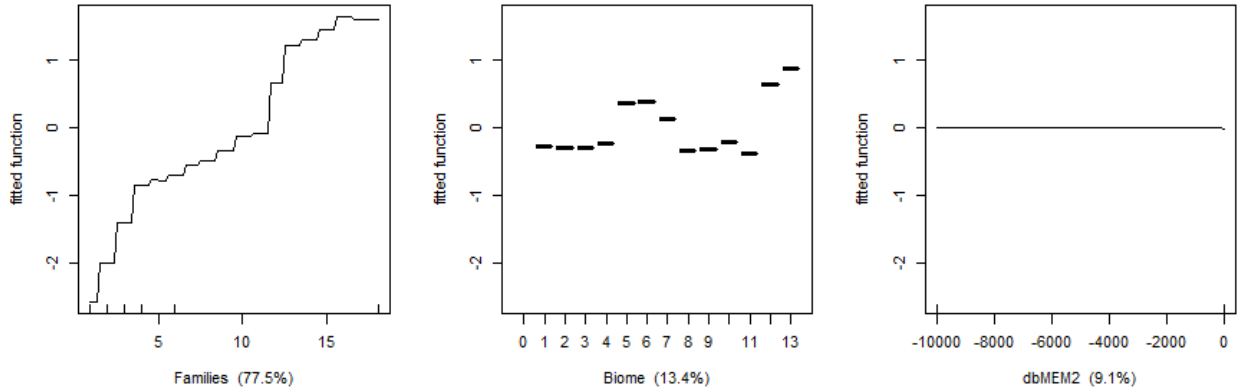

**Supplementary Figure 3.** Boosted regression tree outputs for predicting species richness. Y-axis values represent relative estimates of species richness. Percentages on x-axes indicate contributions of each variable to the overall model. Families = number of amphibian families documented in pixel; biomes (left to right) = tundra permafrost, tundra interfrost, boreal semi-arid, boreal humid, temperate semi-arid, temperate humid, warm Mediterranean, cold Mediterranean, tropical desert, temperate desert, cold desert, tropical semi-arid, and tropical humid; dbMEM2 = spatial eigenvector 2. Notches above x-axes represent partitioning of the pixel data into 10 equal quantiles. As BRTs are prediction-focused models, these outputs should not be interpreted as causal linkages. Predictor variables that are listed in the methods but not shown here were discarded due having a relative influence on the final model totaling < 5% of the most influential variable in the model.

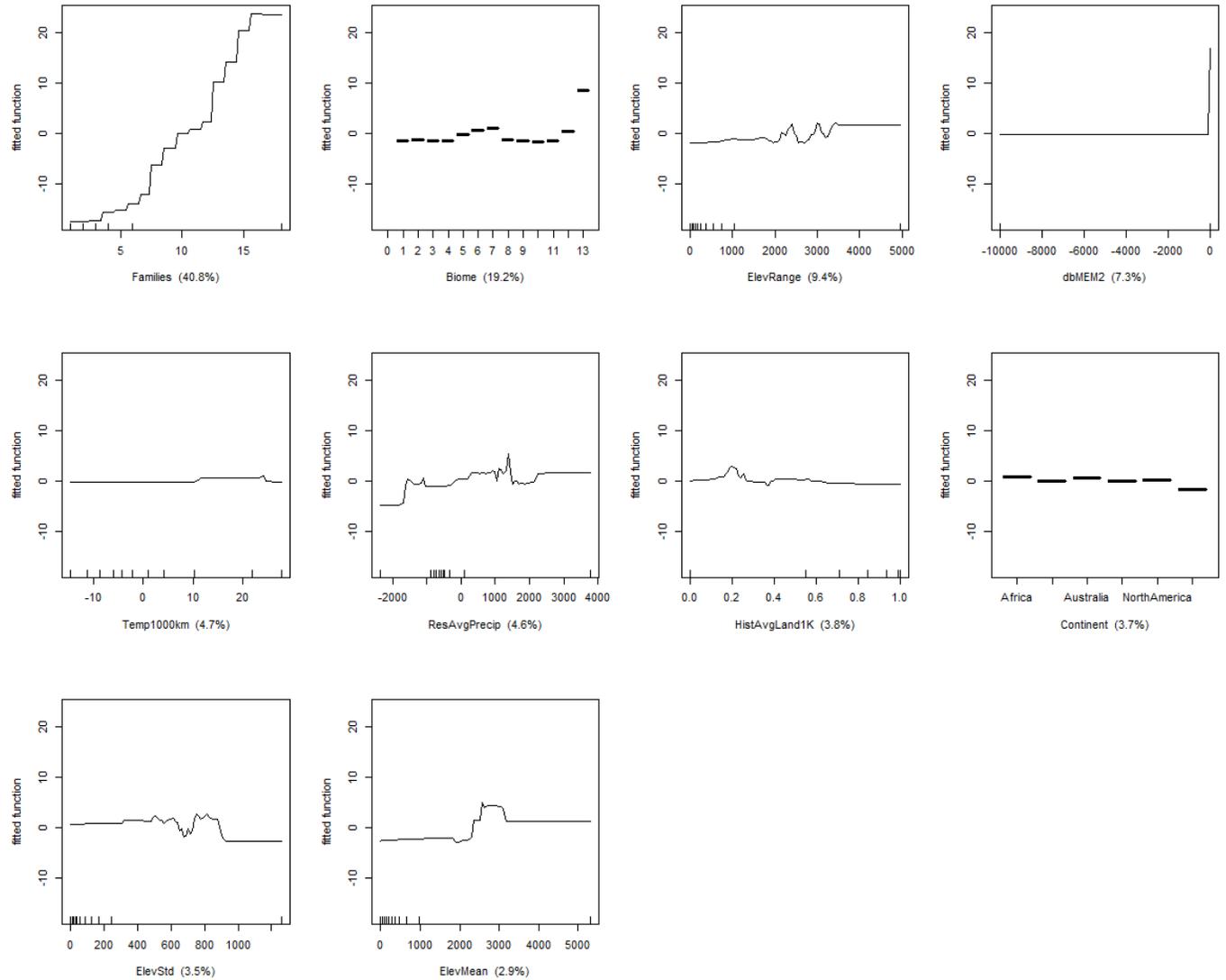

**Supplementary Figure 4.** Boosted regression tree outputs for predicting integrative conservation priority score (Button & Borzee, 2022). Families = number of amphibian families documented; ElevRange = maximum minus minimum elevation; Temp1000km = average annual temperature within 1000km; biomes (left to right) = tundra permafrost, tundra interfrost, boreal semi-arid, boreal humid, temperate semi-arid, temperate humid, warm Mediterranean, cold Mediterranean, tropical desert, temperate desert, cold desert, tropical semi-arid, and tropical humid; continents (left to right) = Africa, Asia, Australia/Oceania, Europe, North America, and South America; dbMEM2 = spatial eigenvector 2; ElevStd = standard deviation of elevation; ResAvgPrecip = average annual precipitation (scaled relative to spatial eigenvector 2; see methods); ElevMean = average elevation; HistLand1K = historical average percent of area within 1000km covered by land. All other details are the same as in Supplementary Figure 3. Predictor variables that are listed in the methods but not shown here were discarded due having a relative influence on the final model totaling < 5% of the most influential variable in the model.

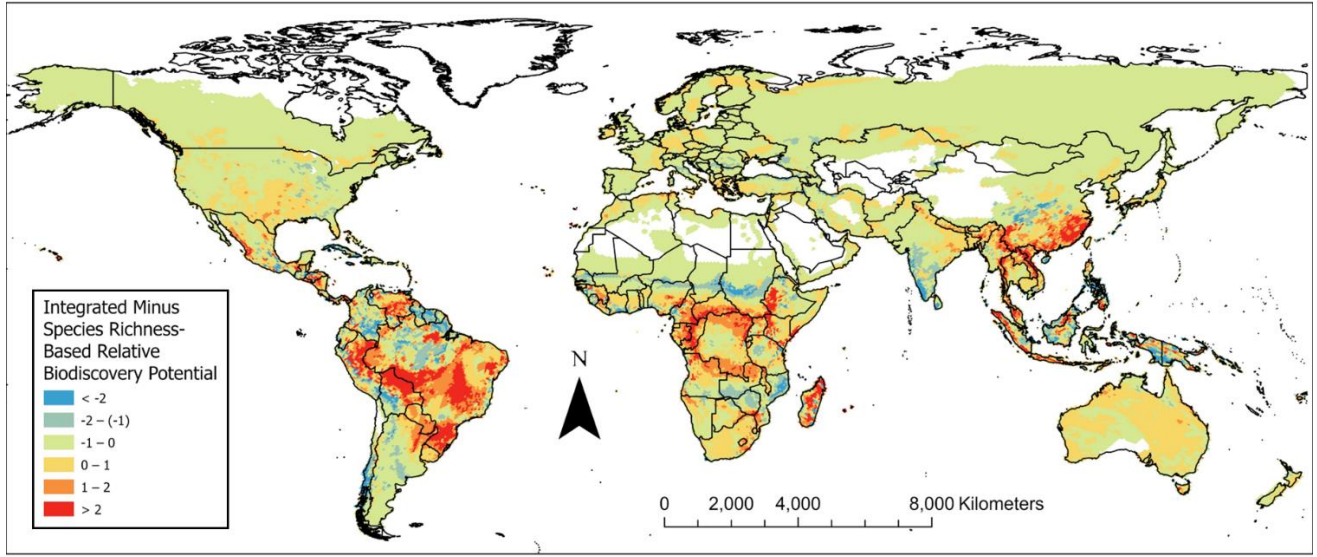

**Supplementary Figure 5.** Differences between relative biodiscovery priorities identified using our species richness versus integrative conservation-based scoring approach.

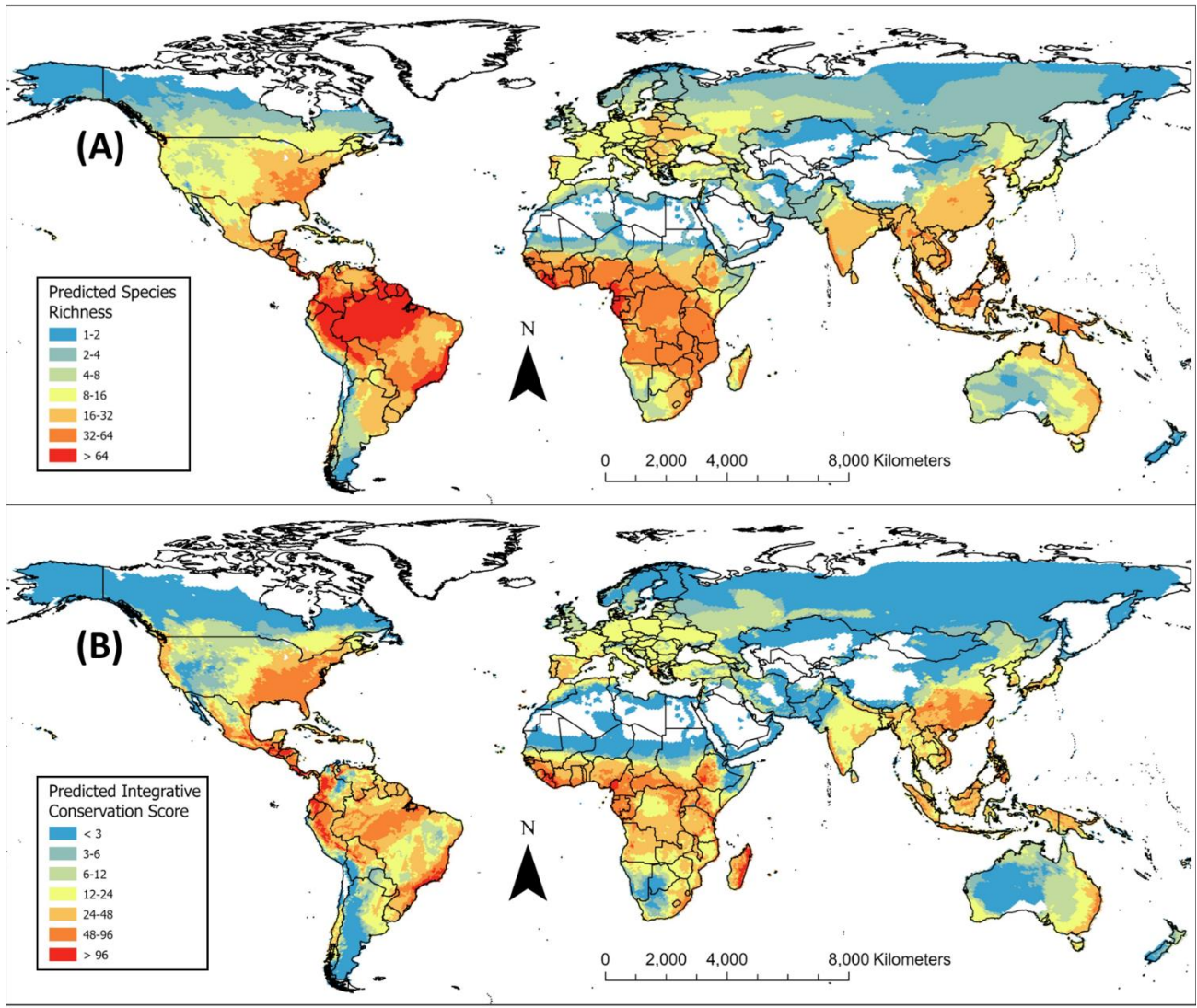

**Supplementary Figure 6.** Model-based predictions of total species richness (A) and total integrative conservation priorities (B) based on BRT model.